\documentclass[aps,prb,reprint,groupedaddress,showpacs]{revtex4-1}
\usepackage{amsmath}
\usepackage{bm}
\usepackage{graphicx}
\usepackage{longtable}
\RequirePackage{ifthen}

\usepackage[usenames,dvipsnames]{color}
\usepackage{ulem} 

\newcommand{\green}{\textcolor{OliveGreen}}

\newcommand{\mf}[1]{\mathbf #1}

\begin{document}

\title{Charge transfer and electron-phonon coupling in monolayer FeSe  on Nb doped SrTiO$_3$}


\author{Yuanjun Zhou}
\affiliation{Department of Physics, Columbia University, New York, New York 10027, USA}
\author{Andrew J. Millis}
\affiliation{Department of Physics, Columbia University, New York, New York 10027, USA}


\date{\today}

\begin{abstract}
Monolayer films of FeSe grown on SrTiO$_3$ substrates are  electron doped relative to bulk and exhibit a significantly higher superconducting transition temperatures. We present density functional calculations and a modified Schottky model incorporating the strong paraelectricity of SrTiO$_3$ which demonstrate that the doping may be due to charge transfer from SrTiO$_3$ impurity bands driven by work function mismatch.    Physically relevant levels of Nb doping are shown to lead to doping of the FeSe compatible with observation.  The coupling of electrons in FeSe to polar phonons in the depletion regime of the SrTiO$_3$ is calculated. A $\lambda\sim 0.4$ is found; the coupling to long-wavelength phonons is found to be dominant. 
\end{abstract}



\maketitle

\def\scr{\scriptsize}
\def\draftversion{false}
\def\draftversion{true}
\ifthenelse{\equal{\draftversion}{true}}{
  \marginparwidth 2.7in
  \marginparsep 0.5in
  \newcounter{comm} 
  \def\commnext{\stepcounter{comm}}
  \def\commtext{{\bf\color{blue}[\arabic{comm}]}}
  \def\commmar{{\bf\color{blue}[\arabic{comm}]}}
  \def\yjm#1{\commnext\marginpar{\small YJZ\commmar: #1}\commtext}
  \def\sm#1{\commnext\marginpar{\scr\green{SAVED\commmar: #1}}\commtext}
  \def\tnewpage{\marginpar{\small Temporary newpage}\newpage}
}{
  \def\yjm#1{}
  \def\sm#1{}
  \def\tnewpage{}
}

\section{Introduction}
The discovery of superconductivity in iron pnictide and calcogenide materials  \cite{Fe-based-first} and the subsequent investigation of the many remarkable properties of these materials \cite{Stewart-review}  has added a new dimension to our understanding of superconductivity. The general consensus has been that the materials demonstrate the existence of a third route to high-T$_c$ superconductivity, based neither on very high phonon frequencies \cite{H3S-203K}  nor on the physics of two dimensional spin $1/2$ carriers near a Mott transition \cite{Anderson1987Science}. Orbital degeneracy leading to a Fermi surface with multiple sheets and to  Hunds metal \cite{Mravlje13}, nematic and stripe \cite{Fernandes-nematic-review} behavior, as well as pairing based on electron-electron interaction-induced scattering of carriers between zone-edge electron pockets and zone-center hole pockets have been regarded as key to the physics \cite{Stewart-review}. 

Recent studies of  monolayer FeSe thin films epitaxially grown on SrTiO$_3$ (STO) substrates challenge this understanding. The transition temperature of bulk FeSe is lower than 10K and the record high transition temperature for bulk members of the Fe-pnictide family is 56K \cite{Fe-bulk-56K}, but in the monolayer films transition temperatures as high as 100K has been reported   \cite{Ge-100K}. Further, monolayer FeSe on STO  is heavily electron doped (about 0.12 $e^-$ per Fe, as inferred from the photoemission-determined Fermi surface \cite{Tan-nmat}) so that the zone-center hole pocket  found in bulk pnictides is fully filled, leaving only the electron pocket at the M point, indicating that the zone face to zone center scattering processes are not crucial to the high transition temperature superconductivity. These finding suggest that our understanding of the Fe-Se materials is incomplete. 

FeSe grown epitaxially on STO is not superconducting if the thickness is larger than 2 unit cells. The thickness dependence and the presence of sidebands in photoemission experiments \cite{el-ph_Shen-FSSTO} has been considered as evidence that proximity to the substrate is important and that fluctuations in the substrate help to enhance the transition temperature. Gate-doped FeSe  flakes up to 10 nm thick and potassium doped multi unit-cell FeSe films also exhibit  transition temperatures between 45 and 50K \cite{FeSe-flake-gating, FeSe-Kdoped}, indicating it is possible that proximity to the substrate is not essential for raising the transition temperature above the bulk value, but is important for raising the transition temperature above the highest value observed in bulk materials. The  longitudinal optic (LO) phonon modes of STO are an attractive mechanism for a proximity effect\cite{el-ph_Shen-FSSTO} because these arise from dipolar fluctuations which produce long ranged electric potentials. The proximity of material to a ferroelectric transition (indeed a signal of FE transition near 45K has been observed in some of the materials with high $T_c$ \cite{ZXShen-FEz_FSSTO}) may enhance the strength of these fluctuations. 

In this paper we investigate monolayer FeSe on STO theoretically. We perform density functional calculations  to obtain the structure and to determine band offsets. The band offsets are found to be large, raising the possibility of charge transfer in the experimentally relevant case where the STO  is lightly electron-doped by Nb impurities or O vacancies.  We treat charge transfer from the impurity bands by adapting the usual  Schottky model to  incorporate the physics of nearly ferroelectric STO. Observed values of electron doping in the monolayer are consistent with charge transfer from impurity bands characteristic of the known  doping levels of the STO substrates. The oxygen vacancy effect noted in Ref. \onlinecite{FeSeSTO-Ovac} may also be relevant. 

A generic consequence of the Schottky model is that the SrTiO$_3$ impurity bands are fully depleted over a many-unit cell wide region near the interface.    Electric fields in this depletion region are not screened, so that polar phonons in this region may couple to the electrons in the FeSe. We determine the electron-phonon coupling Hamiltonian and compute the leading contribution to the electron self energy, which corresponds to a coupling parameter $\lambda \sim 0.4$ (the precise value depends on parameters whose values are not well established), with the coupling dominated by long wavelength phonons.  Our work complements previous studies \cite{XiangYY-2012-el-ph,DHLee-CPB,Rademaker_el-ph_FSSTO}  of the effect of SrTiO$_3$ polar phonons on electrons in FeSe, in which a phenomenological model was used.

\section{First-principles calculations}

This section presents that results of first principles calculations that enable an estimation of the structure and the basic energetics of monolayer FeSe on STO. Our first-principles calculations were performed using the spin-dependent generalized gradient approximation with the Perdew-Burke-Ernzerhof (PBE) parametrization \cite{PBE} as implemented in the {\it Vienna Ab initio Simulation Package} (VASP-5.3\cite{vasp1,vasp2}). Projector augmented wave (PAW) potentials were used,\cite{paw,paw2} with 6 valence electrons for Se ($4s^24p^4$), 8 for Fe ($3d^74s^1$), 10 for Sr ($4s^24p^65s^2$), 10 for Ti ($3p^63d^24s^2$) and 6 for O ($2s^22p^4$).  The energy cutoff was 500eV. A $\sqrt 2\times\sqrt 2$ in-plane cell was used to investigate the consequences of octahedral rotations; as these were found not to be important a $1\times 1$ in-plane cell was used to generate the main band structure results. A $6\times 6\times 1$ k-point mesh and $5\times 10^{-3}$  eV/\AA\, force threshold  were used for structural relaxations.  A $14\times 14\times 1$ k-point mesh was used for density of states calculations.  Phonons of bulk STO are calculated by the frozen phonon method in VASP, using a $3\times 3\times 3$ supercell and $4\times 4\times 4$  k-point mesh, and post-processed  via Phonopy\cite{phonopy}.

\begin{figure}
 \includegraphics[width=\columnwidth]{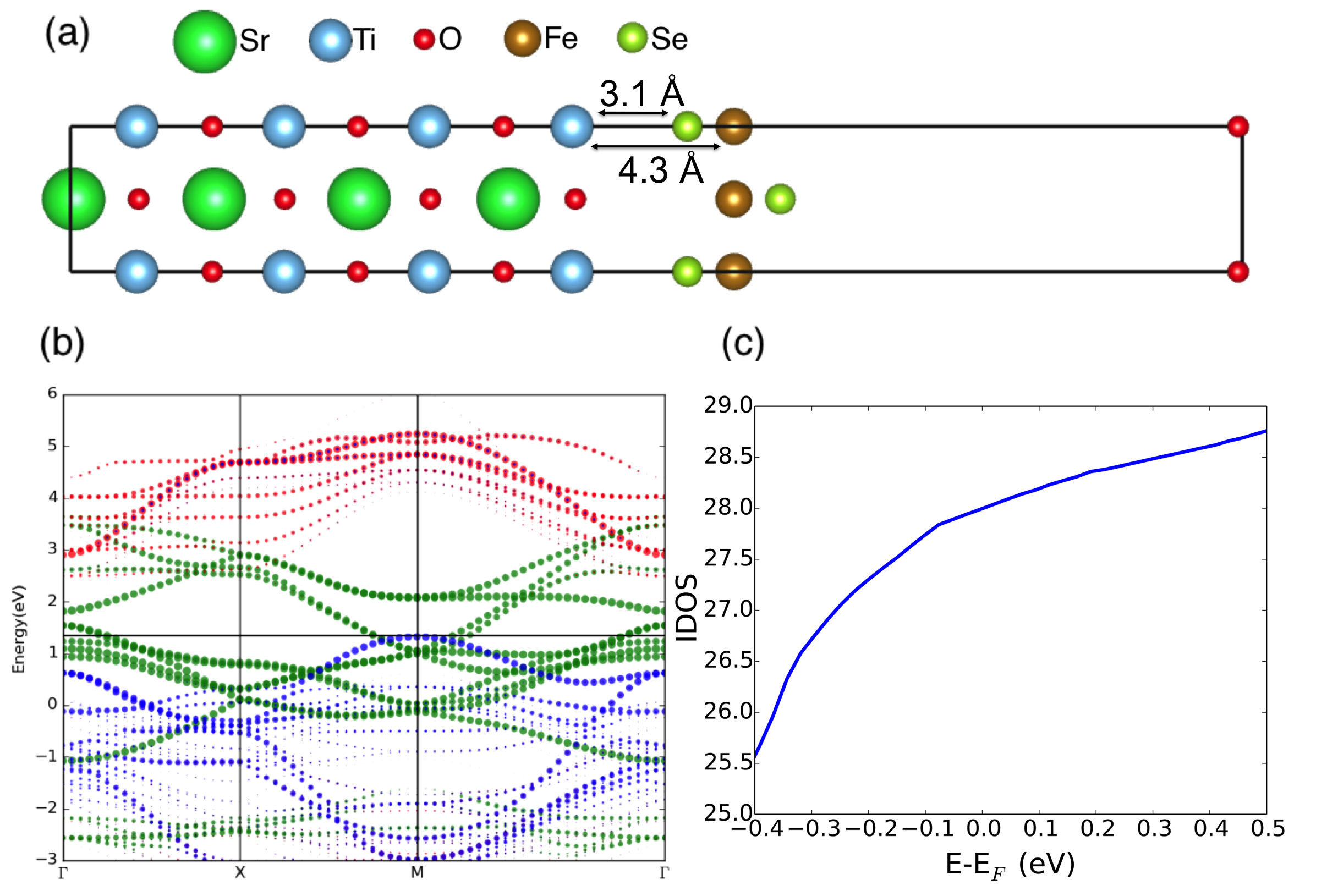}%
 \caption{\label{struct}(a) Side view of relaxed structure I (bottom layer Se aligned with Ti columns, see main text) obtained from calculations that disregard the octahedral rotations of the SrTiO$_3$.  
(b) Energy bands found for the relaxed structure shown in (a). The Fe $d$ bands are in green,  Ti $d$ and O $p$ bands arising from atoms near the STO-FeSe interface are shown  in red and blue respectively.  The vacuum side O bands  are more than 1.5 eV beneath the Fermi level and are not shown.  (c) The integrated density of states per 2 Fe of a freestanding monolayer of  FeSe with atomic positions corresponding to relaxed structure I. 
 }
 \end{figure}

The monolayer of FeSe can be viewed as a layer of  Fe atoms sandwiched between two layers of Se atoms. We assume epitaxial conditions, so the in-plane lattice constant of the FeSe layer is set to be equal to that of STO ($a=3.935$\AA\, in our calculation). Note that there are two Fe atoms per unit cell.  We modeled the substrate as a four unit cell-thick slab of SrTiO$_3$ and included a  vacuum layer with thickness equal to  three unit cells of the SrTiO$_3$ (see panel (a) of Fig.~\ref{struct}). The relative alignment of the FeSe layer with respect to the SrTiO$_3$ is not known. We  considered two configurations:  Structure I, in which the bottom layer Se atoms align with the Ti columns of the SrTiO$_3$ and  Structure II, in which  the bottom layer Se atoms align with the Sr columns of the STO. Structure I is depicted in panel (a) of Fig.~\ref{struct}. 

For both structures, we choose as an initial state the striped antiferromagnetic (AFM) state  that was considered to be the ground state for bulk FeSe \cite{Singh-PRB-FSbands,Xiangtao-FSbulkbands}. Our spin polarized GGA calculations for bulk FeSe (not shown)  also find a striped antiferromagnet ground state. However, for the monolayer we find that the calculation converges to a nonmagnetic state. 

After relaxation of atomic positions we find that structure I is 27 meV/Fe lower in total energy than structure II, and we thus take it to be  the ground state structure in this paper. However the essential features (including for example band offsets) are the same for both structures. We investigated the consequences of the $a^0a^0c^-$ rotation that occurs in bulk SrTiO$_3$ below 105K \cite{STO-Unoki1967} by comparing calculations in which this rotation is allowed and forbidden. We find that the octahedral rotation is suppressed in  the  TiO$_2$ layer closest to the FeSe and that (not shown) the atomic and electronic structure of the FeSe is essentially the same whether or not SrTiO$_3$ octahedral rotations are considered. We therefore disregard the rotations in what follows, and use a $1\times 1$ unit cell in the calculations subsequently reported. Panel (a) of Fig.~\ref{struct} shows the relaxed structure I obtained without including SrTiO$_3$ rotations. The proximal Se is found to be separated from the surface Ti by 3.1\AA, and the distance to the Fe plane is 4.3\AA, in agreement with previous DFT calculations\cite{XiangTao-FSSTObands}. 
 The Fe-Se-Fe bond angle is found to be 114.82$^\circ$, larger than the FeSe bulk value\cite{Yin-NatMat2011}. The difference is  due in part  to the tensile epitaxial strain ($\sim 4\%$) generated by the STO and in part may arise from the difference in magnetic state: in bulk calculations the non-magnetic state  has a larger bond angle that the antiferromagnetic state \cite{Yin-NatMat2011,K2FeSe-bondangle} 

Panel (b) of Fig.~\ref{struct} shows the important energy bands computed for relaxed structure I.   The FeSe layer is metallic; its chemical potential sets the Fermi level which is slightly above the top of the SrTiO$_3$ valence band.  The FeSe fermi level (horizontal solid line at $E = 1.34$eV) lies in the SrTiO$_3$ energy gap so in this calculation there is no charge transfer between the FeSe and the SrTiO$_3$. As is also the case for  stoichiometric bulk pnictides,  the monolayer is a compensated semimetal with two  zone center hole pockets and two zone-face  electron pockets, centered at the $M$ $(\pi/a,\pi/a)$ 
(using the notation appropriate  to the Brillouin zone corresponding to the two Fe unit cell). 

Panel (c) of Fig.~\ref{struct} presents the integrated density of states of an isolated FeSe layer in the near fermi surface region. The break in slope visible at $E-E_F\approx 0.2$eV is due to the filling of the $\Gamma$ point hole pocket. The slope for energies greater than this value gives the density of states of the electron pockets $\mathcal{N}(\varepsilon)\approx 0.66$ states/eV per Fe. Modeling the two electron pockets as ellipses with masses $m_1$ and $m_2$ in the principal directions we conclude that if $b$ is the Fe-Fe distance then  $\sqrt{m_1m_2}b^2 / \hbar^2 \approx 1$ state/eV. 

Angle-resolved photoemission experiments on monolayer FeSe find that the hole pocket is fully filled but also find that the top of the hole pocket lies below the bottom of the electron pocket  \cite{Tan-nmat}.  However, experiment and DFT calculations find that bulk FeSe is a compensated semimetal, in which necessarily the top of the hole band lies above  the bottom of the electron bands, with the Fermi level in between. This suggests that a rigid band approximation for the effects of doping is inadequate, and that there is a density dependent change in the band structure which as electrons are added acts to lower the energy of the top of the hole bands relative to the bottom of the electron band. Deviations from the rigid band picture may explain why our DFT results indicate that $Q\approx 0.17e/Fe$ must be added to the monolayer to fill the hole band, while in experiment a $Q\approx 0.12$ is found.

\begin{figure}
 \includegraphics[width=\columnwidth]{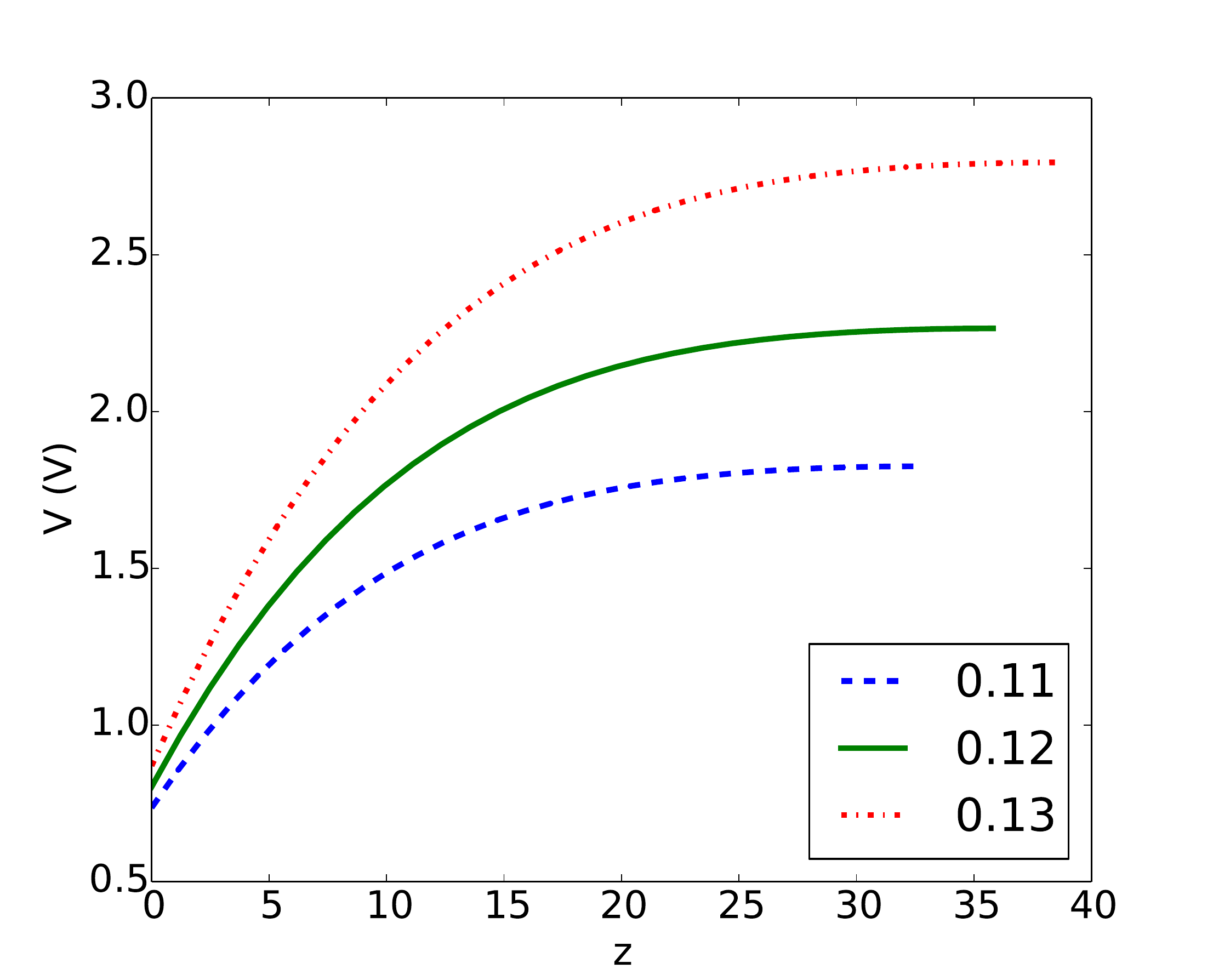}%
 \caption{\label{schottky} Charge transfer at the FeSe-Nb:SrTiO$_3$ interface. Spatial variation of potential computed from Eq.~\ref{VinSTO} with $N_d=1.1\times10^{20}/cm^3$ and $Q=0.13/Fe$ (upper curve), $0.12$ (middle) and $0.11$ (lowest) with other parameters as given in text.  $y$ axis: potential in eV; $x$ axis distance from interface in units of STO pseudocubic lattice constant.}
 \end{figure}

\section{Charge transfer}

The first principles calculations reported in the previous section assume ideal SrTiO$_3$, but the SrTiO$_3$ substrates used in experiments are lightly electron-doped. The electrons reside in impurity bands which are typically close to the bottom of the SrTiO$_3$ conduction band, while the FeSe fermi level is close to the top of the SrTiO$_3$ valence band, suggesting a work function difference  slightly less than the SrTiO$_3$ band gap ($\sim 2eV$ in DFT band  calculations and $\sim 3eV$ in experiment) which may drive charge transfer.  In this section we develop a Schottky model modified to take into account the strong field dependence of the dielectric constant of SrTiO$_3$, to analyse this physics. 

The basic situation is that charge doped into the FeSe  produces an electric field, which bends the STO bands, so that within a depletion regime extending a distance $d$ of the interface  the electrochemical potential is below the bottom of the  STO impurity bands so there are no carriers.  The full depletion region is followed by a partial depletion region over which  the potential changes from the bottom of the STO impurity band to the Fermi level of the bulk STO impurity band. The width of the partial depletion region is set by the Thomas-Fermi screening length $\lambda_{\text{TF}}$  but we shall see that the details of this region are not important for the considerations presented here.    

We model the carriers doped into the FeSe as a sheet of areal charge density $-eQ$, set back a distance $d_{\text{Fe}}\approx 4.3$\AA\, from the interface. Following Stengel \cite{Stengel-STO-dielectric} we model the field in the  STO as a macroscopic displacement field $D$ depending only on distance $z$ from the interface, which we take to be at $z=0$. In the full depletion region $D$ obeys the equation
\begin{equation}
\frac{dD(z)}{dz}=4\pi eN_d
\label{Deq}
\end{equation}
with $N_d$ (typically $\sim 10^{20}$)  the volume density of dopant ions. Solving Eq.~\ref{Deq} we find
\begin{equation}
D(z)=D_0\left(1-\frac{z}{z_0}\right)~~(0\leq z \leq z_0)
\end{equation}
and $D(z>z_0)=0$.

The value $D_0$ of the displacement field at $z=0$ is set by the charge  in the FeSe as
\begin{equation}
D_0=-4\pi e Q
\label{D0}
\end{equation}
and the charge compensation length $z_0$ is defined as the ratio of the FeSe charge to the  volume density of dopant ions  $N_d$ 
\begin{equation}
z_0=\frac{Q}{N_d}
\end{equation}
For typical values $Q=0.125 e$/Fe and $N_d=10^{20}/cm^3$ $z_0\approx 15nm\sim 40$ unit cells. The size $d$ of the full depletion region is of course somewhat less than $z_0$, and for $z>d$ the free charge of the SrTiO$_3$ impurity bands must be included in Eq.~\ref{Deq}.

To compute the potential we determine the electric field $E=D/\epsilon$ and integrate. Thus we need to know the dielectric constant in the presence of displacement fields of the order of the value $D_0 \sim 3.\times 10^8V/cm$ produced by the charge density $Q=0.125$/Fe.
We may assume the dielectric constant in the FeSe region takes a constant value $\epsilon_\text{FeSe} = 15$ independent of $D$\cite{STO-Born-charge}, but the nearly ferroelectric nature of SrTiO$_3$ means that the dielectric constant in this material is large and field-dependent \cite{STO-field-permittivity}.   The displacement fields  ($D_0 \sim 10^8V/cm$)  imply electric fields very much larger than those for which $\epsilon_\text{STO}$ has been measured \cite{STO-field-permittivity}, so theoretical input is required. We use the calculations of Stengel \cite{Stengel-STO-dielectric} which indicate that $\epsilon$ depends strongly on $D$ but not strongly on its gradients and  is well described as a Lorentzian 

\begin{equation}
\epsilon(D)=\epsilon_\infty+\frac{\epsilon_0-\epsilon_\infty}{1+\left(\frac{D}{D^\star}\right)^2}
\label{epsilonofD}
\end{equation} 
with $\epsilon_\infty = 5.18$, $\epsilon_0 = 500$ and $D^*$ the displacement produced by a sheet charge of density $\approx 0.05e$/Fe.

Relative to the FeSe Fermi level, the potential at depth $z<d$ in the SrTiO$_3$ is then
\begin{equation}
V(z)=\frac{D_0d_{\text{Fe}} }{\epsilon_\text{FeSe}}+\int_0^zdz^\prime\frac{D(z^\prime)}{\epsilon(D(z^\prime))}
\label{VinSTO}
\end{equation}

We are interested in potential shifts of the order of the SrTiO$_3$ band gap $\sim 3eV$. In principle, we must also account for the shift in the FeSe Fermi level as carriers are added. The monolayer FeSe integrated density of states is shown in Fig.~\ref{struct}(c).  We see that the change in FeSe fermi level corresponding to a $Q\sim 0.2/Fe$ is small, of order a few tenths of an eV. Also, as shown in the appendix, we may neglect the details of the behavior in the Thomas-Fermi regime (over which the potential changes by an amount of the order of the impurity band width, also $\sim 0.1eV$), and simply compute the potential profile caused by a given level of charge in the FeSe.

Fig.~\ref{schottky} shows representative results. We see that the potential rises rapidly and in a nearly linear manner for the first $\sim 10$ unit cells. This  pseudolinear behavior comes from the increase in $\epsilon$ as the charge is progressively screened and $D$ decreases. Following the pseudolinear regime, the potential then slowly approaches its saturated value over the last several unit cells.  The value of linear response dielectric constant $\epsilon_0$ affects only  the final asymptotic  approach to the final value, and for the same reason the properties of the partly depleted region do not significantly affect the magnitude of the induced charge. The sensitivity of the results to $Q$ implies that uncertainties in the values of $\epsilon_\infty$ and $N_d$ do not have an important effect on the final charge density.  For example, changing $\epsilon_\infty$ to 2 or 15 and $N_d$ to $1.5\times10^{20}/cm^{3}$ produces essentially the same results. Thus  our results are  sensitive to $D^\star$ and, to a less extent, the high-D limit $\epsilon_\infty$. The results are only weakly sensitive to the zero-field value $\epsilon_0$, as long as it is large. Electron doping of the order of the values seen in experiment may thus reasonably be understood as coming from charge transfer from the SrTiO$_3$ impurity bands.

\section{Coupling of SrTiO$_3$ Dipolar Phonon Modes to FeSe electrons}

Electric fields produced by  polar phonons in the depletion region are not screened, and so produce potential fluctuations that can couple to electrons in the FeSe layer. This type of electron-phonon coupling has been discussed in the context of monolayer FeSe on SrTiO$_3$ \cite{el-ph_Shen-FSSTO}, but in that work the interaction was taken to be local, arising from the dipoles nearest to the interface. Here we present a general treatment. 

We assume that in each unit cell of the depletion region there is a dipole, characterized by position $\mf R$, displacement $\mf u$ and effective charge $eZ_d$. The potential at position $\mf r$ in the FeSe layer due to  the SrTiO$_3$ dipoles is 
\begin{equation}
V(\mf r)=\frac{eZ_d}{\epsilon_\infty}\sum_{\mf R}\frac{\mf u_R\cdot\left(\mf r+\hat{z}d_{Fe}-\mf R\right)}{\left|\mf r+\hat{z}d_{Fe}-\mf R\right|^3}
\label{Vdipoleplane}
\end{equation}
where $\epsilon_\infty$  is the dielectric constant that  SrTiO$_3$ would have in the absence of the dipoles.  Here the sum over $\mathbf{R}$ runs over the depletion and Thomas-Fermi regions; the fields of dipoles farther inside the SrTiO$_3$ are screened. 

The corresponding electron-phonon term in the Hamiltonian is then
\begin{equation}
H_{ep}=\sum_{\mf q}\rho_{\mf q}eV(-\mf q)
\label{Hep}
\end{equation}
where $\rho_q$ is the electron number density operator at momentum $\mf q$. By Fourier transforming Eq.~\ref{Vdipoleplane} with respect to the in-plane coordinate and  noting that local field corrections  are of order $e^{-2\pi d_{Fe}/a}\sim 10^{-3}$ so can be neglected we obtain
\begin{equation}
eV(q)=\frac{2\pi e^2Z_de^{-qd_{Fe}}}{\epsilon_\infty a^2}\sum_{Z_J\leq z_0}e^{-qZ_J}\mf u_{q,J}\cdot\left(i\hat{q}+\hat{z}\right)
\label{Vdipoleplanefinal}
\end{equation}
where $q=|\mf q|$, $J=1,2,...$ labels the Ti planes in the depletion layer, $Z_J=a\left(J-\frac{1}{2}\right)$ is the distance of the J$^\text{th}$ Ti plane from the interface and we have included the effects of screening of dipoles deep in the SrTiO$_3$ by a screening length $q_0^{-1} \approx z_0$. Thus the electron-phonon coupling extends over a number of planes determined by the smaller of the inverse of the in-plane momentum $q$ and the approximate thickness, $z_0$, of the depletion layer.

We now quantize the dipole fluctuations.  For simplicity we assume that the dipole fluctuations in the depletion region are  those of undoped SrTiO$_3$ and we take in to account the interface by assuming that the z-derivative of the phonon field  vanishes at the interface. First-principles calculations \cite{STO-Born-charge} indicate that SrTiO$_3$ has three important dipole-active (LO) phonon modes. We calculated the frequencies, effective charges, and masses of these modes. Results (fully consistent with those of Ref. ~\onlinecite{STO-Born-charge})  are given in Table ~\ref{table.ep5.18}. 
\begin{table}[ht]
 \caption{ \label{table.ep5.18}$\Gamma$ point STO LO modes and mode-dependent effective charges for $\epsilon_\infty = 5.18$. The effective masses are in the unit of the mass of a proton.}
  \begin{ruledtabular}
   \begin{tabular}{ c c c c}
    LO\# & $\Delta$(meV) &$Z$  &$M(m_p)$ \\ 
    \hline 
    1 & 18.3  & 0.4 & 46.83\\
    2 & 56.0  & 2.0 & 17.87\\
    3 & 97.1 & 8.0 & 18.96\\
  \end{tabular}
  \end{ruledtabular}
\end{table}

We then write the total dipole as a sum of fluctuations of these three modes, 

\begin{equation}
\begin{split}
\mf u(\mf q,J)=&\sum_\alpha\int_0^{2\pi/a} \frac{dq_za}{2\pi}e^{iq_zZ_J}\left(\mf u_\alpha(\mf q,q_z)+\mf u_\alpha(\mf q,-q_z)\right) \\
=&
a\sum_\alpha\int_0^{2\pi/a} \frac{dq_za}{2\pi}e^{iq_zZ_J}\sqrt{\frac{\hbar}{2M_\alpha\omega^L_{\alpha\mf q,q_z}}} \\
&\times \bigg(\hat {u}(\mf q,q_z)\left(b^\dagger_{\alpha\mf q,q_z}+b_{\alpha-\mf q,-q_z}\right) \\
& +\hat {u}(\mf q,-q_z)\left(b^\dagger_{\alpha\mf q,-q_z}+b_{\alpha-\mf q,q_z}\right) \bigg)
\label{phonons}
\end{split}
\end{equation}
Here $b^\dagger_{\alpha q,q_z}$ creates a longitudinal polar phonon of mode $\alpha$ with in-plane momentum $q$ and out of plane momentum $q_z$; $\hat{u}$ is the  phonon polarization and $\omega_\alpha^L$ is the phonon frequency.

Combining the formulas and summing over $J$ yields

\begin{equation}
eV(\mf q)=\sum_\alpha\int\frac{dq_za}{2\pi}g^\alpha_{\mf q,q_z}\left(b^\dagger_{\alpha\mf q,q_z}+b_{\alpha-\mf q,-q_z}\right)
\label{Vfinal}
\end{equation}
with coupling function $g^\alpha$ between the electrons and mode $\alpha$ given by
\begin{eqnarray}
g^\alpha_{\mf q,q_z}&=&\frac{2\pi e^2Z^\alpha}{\epsilon_\infty a^2} \sqrt{\frac{\hbar^2}{2M_\alpha\omega^L_{\alpha,q,q_z}}}
\frac{e^{-\left(q-iq_z\right)\left(d_{Fe}+\frac{a}{2}\right)}}
{1-e^{-a\left(q+q_0-iq_z\right)}}
\label{gdef}
\\
&&\approx \frac{2\pi e^2Z^\alpha}{\epsilon_\infty a^3 }
\sqrt{\frac{\hbar^2}{2M_\alpha\omega^L_{\alpha,q,q_z}}}\frac{1}{q+q_0-iq_z}
\label{gapprox}
\end{eqnarray}
The second, approximate equality is valid at small $q,q_z$. We see that the interaction is strongest at small $q$ but is cut off when $q$ becomes of the order of the inverse of the size of the depletion layer. 

\begin{figure}
 \includegraphics[width=\columnwidth]{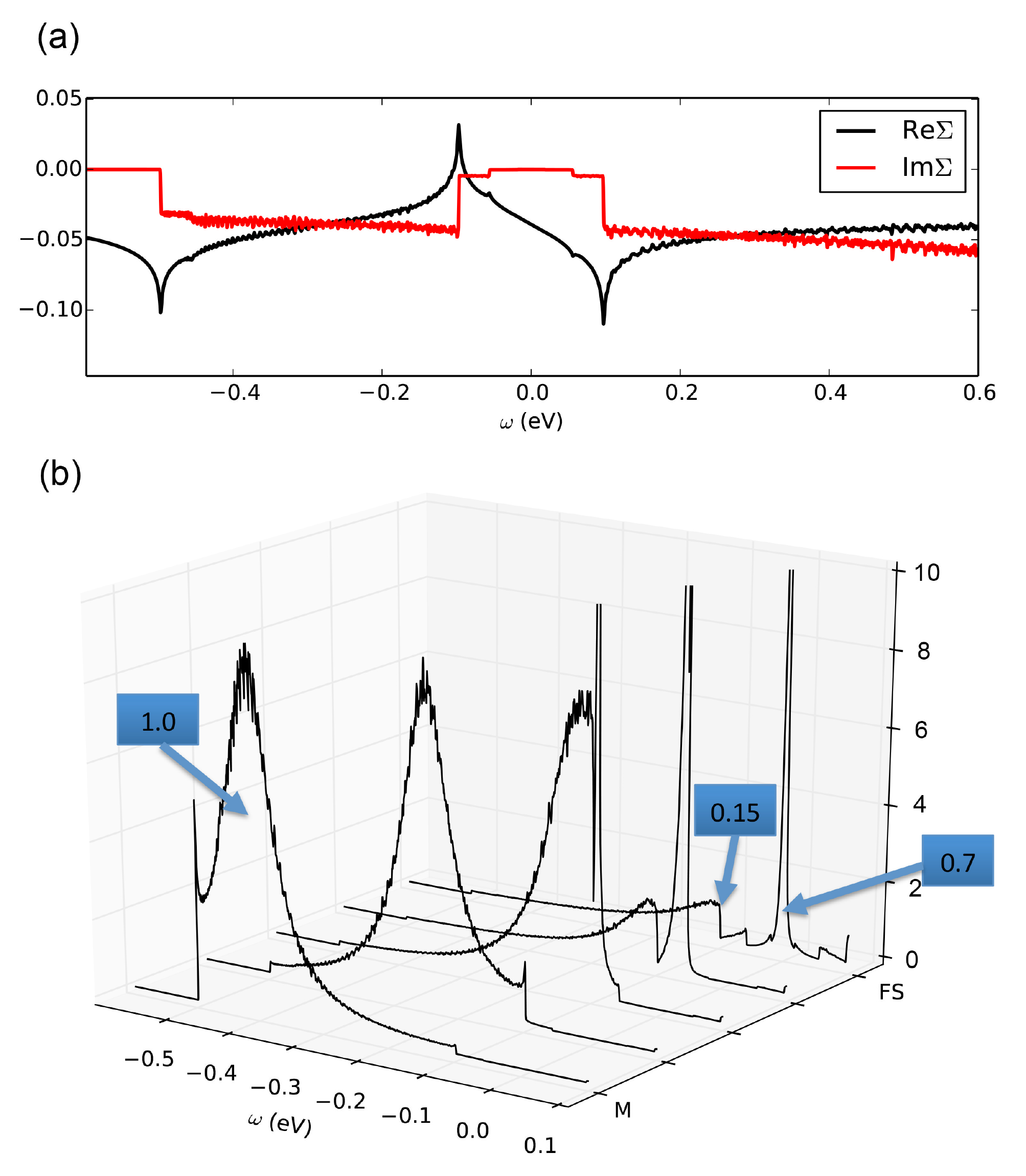}%
 \caption{\label{Sigma} (a) The lowest order electron-phonon self energy at the Fermi surface $k=(0.123,0.123)\frac{2\pi}{a}$ from ($\pi/a,\pi/a$), $4\times 10^6$ equally spaced k points are used for the 2D integration of the self energy in a circular area with the radius $6\pi/a$.  $\omega = 0$ is the Fermi energy. 
 (b) The corresponding spectral functions for different $k$ points from $M$ to the Fermi surface(FS). The integrated areas of peaks under the curves are denoted in blue boxes. Note the total area including both positive (not shown) and negative frequencies is 1.
}
  \end{figure}

A detailed analysis of the consequences of Eq.~\ref{Vfinal} and ~\ref{gdef} will be given elsewhere. Here we present approximate considerations that provide a rough estimate of the electron-phonon coupling strength and the physics of the coupling. The leading order expression for the electron self energy is shown in the upper panel of Fig. ~\ref{Sigma}. The corresponding analytic expression is 

\begin{equation}
\Sigma(\mf k, i\omega) = -T\sum_{\alpha\mf q,q_z,\Omega}|g^\alpha_{\mf q,q_z}|^2 
 D^\alpha(\mf q,q_z, i\Omega) G(\mf k-\mf q, i\omega-i\Omega)
\label{Sigmalowest}
\end{equation}
where the bare phonon and electron propagators are respectively $D^\alpha(q,q_z,\Omega)=2\omega_{\alpha,q}^L/(\Omega^2+(\omega_{\alpha,q}^L)^2)$ and $G(k,\omega)=1/(\omega-\varepsilon_k+\mu)$ and $T$ is the temperature. Using the small momentum form of Eq.~\ref{gapprox}, approximating the $q_z$ integral as determined by the pole in $g$ at $q_z=iq$, assuming $\omega^L=\Delta_\alpha+C(q^2+q_z^2)$, performing the frequency sum as usual  and analytically continuing the fermion frequency to obtain the retarded self energy gives 
\begin{equation}
\label{Sigma_calc}
\begin{split}
\Sigma(\mf k, \omega) =& \sum_\alpha\int \frac{ad^2k^\prime }{(2\pi)^2} \frac{A^2_\alpha}{2\left(\left|\mf k-\mf k^\prime\right|+q_0\right)\Delta_\alpha}\times \\
 &\bigg( \frac{1-f_{k^\prime}+n_{\left|\mf k-\mf k^\prime\right|}}{\omega - \varepsilon_{k^\prime} - \Delta_\alpha +i\delta} + \frac{f_{k^\prime}+n_{\left|\mf k-\mf k^\prime\right|}}{\omega - \varepsilon_{k^\prime} +\Delta_\alpha +i\delta } \bigg) 
\end{split}
\end{equation}
with
\begin{equation}
A^2_\alpha=\left(\frac{2\pi Z^\alpha e^2}{
\epsilon_\infty a}\right)^2\frac{\hbar^2}{2M_\alpha a^2}
\end{equation}
and $f$ and $n$ the Fermi and Bose distribution functions.

For the purposes of estimating the electron-phonon coupling we approximate the electronic energy contours as circles centered on the $M$ points, with dispersion
\begin{equation}
\varepsilon_k=\frac{\hbar^2 k^2}{2m}
\label{dispersionnew}
\end{equation}
k is measured relative to $M~(\pi/a,\pi/a)$ and  effective mass $m$ the geometric mean of the masses corresponding to the dispersion along the two principal axes of the ellipse;  $m=\sqrt{m_1m_2}$.  Our DFT calculations give   $\hbar^2/2mb^2=0.5$eV with $b$ the Fe-Fe distance while the observed charge $Q=0.12/$Fe yields $k_Fb\approx 0.61$ (recall there are two bands). 

We may now estimate the mass enhancement parameter $\lambda=-\partial \Sigma(k_F,\omega)/\partial\omega |_{\omega= 0}$ at $T=0$, using $d\varepsilon_{k'} = \hbar^2 k'dk'/m$,
\begin{equation}
\lambda=\sum_\alpha \frac{amA^2_\alpha}{8\pi^2\hbar^2\Delta_\alpha} \int d\varepsilon_{k'} I(k,k') 
\frac{1}{(|\varepsilon_{k'}| - \Delta_\alpha)^2}
\label{lambda}
\end{equation}
with 
\begin{equation}
I(k,k^\prime)=\int_0^{2\pi} \frac{d\theta}{\sqrt{k^2+k'^2-2kk^\prime \cos\theta}+q_0}
\label{Idef}
\end{equation}
The $\varepsilon_{k'}$ integral is dominated by $\varepsilon_{k'}$ within $\Delta_\alpha$ of the Fermi surface, i.e. by $k'$ within $\delta k = \Delta_\alpha/\hbar^2v_F$ of $k_F$.
For $k$, $k' \approx k_F$ the integral of $\theta$, $I(k,k')$, is logarithmically divergent, with the log cut off by the larger of $\delta k$ and $q_0$ so that
\begin{equation}
\lambda = \sum_\alpha \frac{amA^2_\alpha}{2\pi^2\hbar^2\Delta^2_\alpha k_F} \ln\frac{\pi  }{ \frac{\Delta_\alpha}{\hbar^2 v_F k_F} + \frac{q_0}{k_F}}
\end{equation}
Evaluating with the effective mass, $k_F$ and LO mode frequencies gives us a total value $\lambda \approx 0.4$ \cite{lambda-notes}.

Fig.~\ref{Sigma} shows the real and imaginary parts of the self energy evaluated from Eq.~\ref{Sigma_calc} using a rough fit to the DFT dispersions over the whole zone with the rigid shift of the Fermi level reflecting the doping effect,
\begin{equation}
 \varepsilon(\mf k) =- t\left(\cos(k_x a)+\cos(k_y a)\right) +\mu,
\end{equation}
centered at $\Gamma$, with the Fermi energy at zero, $t=0.7$ eV and $\mu=1 eV$, 
$a=3.935$\AA\, is the Ti-Ti distance. 

There are uncertainties in the size of the electron-phonon coupling. The basic coupling scales are quadratically sensitive to the background dielectric constant and the phonon frequency (which may be affected by the boundary), and the electronic dispersion is somewhat uncertain. 

Fig.~\ref{Sigma} shows the lowest order self energy near the Fermi surface, and the spectral functions due to the electron-phonon coupling based on above assumptions corresponding to the points at and below the Fermi surface. 
The self energy weakly depends on $k$, the dominant contribution comes from the LO mode with frequency 97.1 meV,  and so the imaginary part of the self energy shows a $2\Delta\approx 200$meV gap near the Fermi level. The other two LO modes generate moderate corrections that are revealed by the small plateaus in the $2\Delta$ gap.
Accordingly, for the case near the Fermi surface, the spectral function shows a delta peak in the  $2\Delta$ gap.
The largest side peaks emerge at about 100 meV away from the delta peak, two other smaller peaks lead by the two minor LO modes can also be found in the $2\Delta$ gap.
Away from the Fermi surface, the quasiparticle peaks get broadened due to the nonzero imaginary part of self energy.

\section{Summary}
The remarkable discovery of unusually high transition temperature superconductivity in monolayers of FeSe on SrTiO$_3$ \cite{Ge-100K} raises many interesting questions. In this paper, we have theoretically addressed two of these, showing that work-function-mismatch-driven charge transfer from the impurity bands of Nb (or O-vacancy) doped STO can account for the doping of the FeSe, that a wide depletion region exists in the STO,  and that polar phonon modes in the associated depletion regime of the STO can couple to electrons in the FeSe layer via a coupling that is not small and  turns out to be divergent at long wavelengths. 
Determining the implications for the superconductivity is an interesting open question. The pnictides are strongly correlated materials, and the interplay between electron-phonon coupling and strong correlations is an unsolved problem. In particular, our estimates of the effect of the phonons on the electronic properties are based on DFT values for the electronic dispersion and compressibility. Both of these may be renormalized by the strong correlations believed to be important in the pnictides.
Investigation for example using the combination of density functional and dynamical mean field theory, would be fruitful. We observe that because the phonons couple to the total charge density the methods of Ref.~\onlinecite{Werner07} are applicable. 

The phonons found to be important in this paper are LO phonons with the relatively high frequency of 0.1 eV, suggesting a large contribution to superconductivity. On the other hand, they are long wavelength fluctuations, and forward scattering typically does not make the dominant contribution to the pairing interaction. This issue as well as the interplay with nematic fluctuations need further investigations.  Connections to related experimental systems also require investigations. Superconductivity with critical temperatures of the order of 50-60K has been observed in FeSe flakes on SiO$_2$/Si surface doped via a liquid electrolyte \cite{FeSe-flake-gating}
and for monolayers of FeSe grown on anatase TiO$_2$ \cite{Ding-FeSeTiO2}. 
The 50-60K transition  temperature is  much higher than the $\sim 8K$ T$_c$  of bulk FeSe but is lower than the 100K reported for monolayer FeSe on SrTiO$_3$. Whether the SiO$_2$ and TiO$_2$ sustain polar phonon fluctuations that have a similar effect to those of SrTiO$_3$ are open questions that require further investigation. Also there  are indications in the liquid electrolyte doped system  that the doping extends more than one or two layers. Any polar fluctuations from the substrate would be screened within one or two unit cells, suggesting that it may be possible to distinguish the effects of the phonons we have calculated here from other physics operating in the system. 
In the end, we find that the charge transfer and the associated electron-phonon coupling are general for the heterostructure of superconductor thin films on top of dielectrics when the work function mismatch is large enough, and it would be interesting to investigate these effects in other systems such as other iron pnictides, chalcoginides and cuprates for which layered structures are available.

\section{Acknowledgement}
We thank K. M. Rabe for helpful discussions.
YZ  is supported by National Science Foundation
under grant No. DMR-1120296. AJM  is supported
by the Department of Energy under grant ER-046169.

\appendix
\renewcommand{\thefigure}{A\arabic{figure}}

\setcounter{figure}{0}

\section{The partial depleted region}
Instead of assuming an abrupt boundary of depletion region in the Schottky model, we consider a smooth transition.
Let us consider a flat impurity band with the band width $W$. 
We set up the coordinates as following, $z$ is normal to the interface with the positive direction points in FeSe, and the surface of STO has $z=0$, so that for $z=-\infty$ the Fermi level sits at the center of the impurity band, corresponding to the doping level $DW/2 = N_d$, where $D$ is the constant density of states of the impurity band and $N_d$ is the electron density in bulk Nb-doped STO.

As in the Schottky model, the impurity band bends along the bottom of the conduction band at the interface. 
We set a critical distance $z_0$ ($z_0<0$) such that when $z<z_0$, there are excessed electrons in STO, but when $z>z_0$ all the excessed electrons are depleted into FeSe.
\begin{equation}
\left\{
\begin{split}
\delta n =& -\frac{2e V N_d}{W} \quad (z<z_0) \\
\delta n = & N_d/2 \quad (z>z_0)
\end{split}
\right .
\end{equation}
where $e =1.6\times 10^{-19}$ C.
Following the Poisson's equation $-\partial^2 V(z) / \partial z^2 = -4\pi e\delta n/\epsilon$, 
for $z>z_0$, we have discussed in the main text.
For $z<z_0$,
\begin{equation}
 \frac{\partial^2 V}{\partial z^2} - \frac{8\pi N_de^2 V}{\epsilon W} =0 
\end{equation}

\begin{figure}[t]
\begin{center}
 \includegraphics[width=0.8\columnwidth]{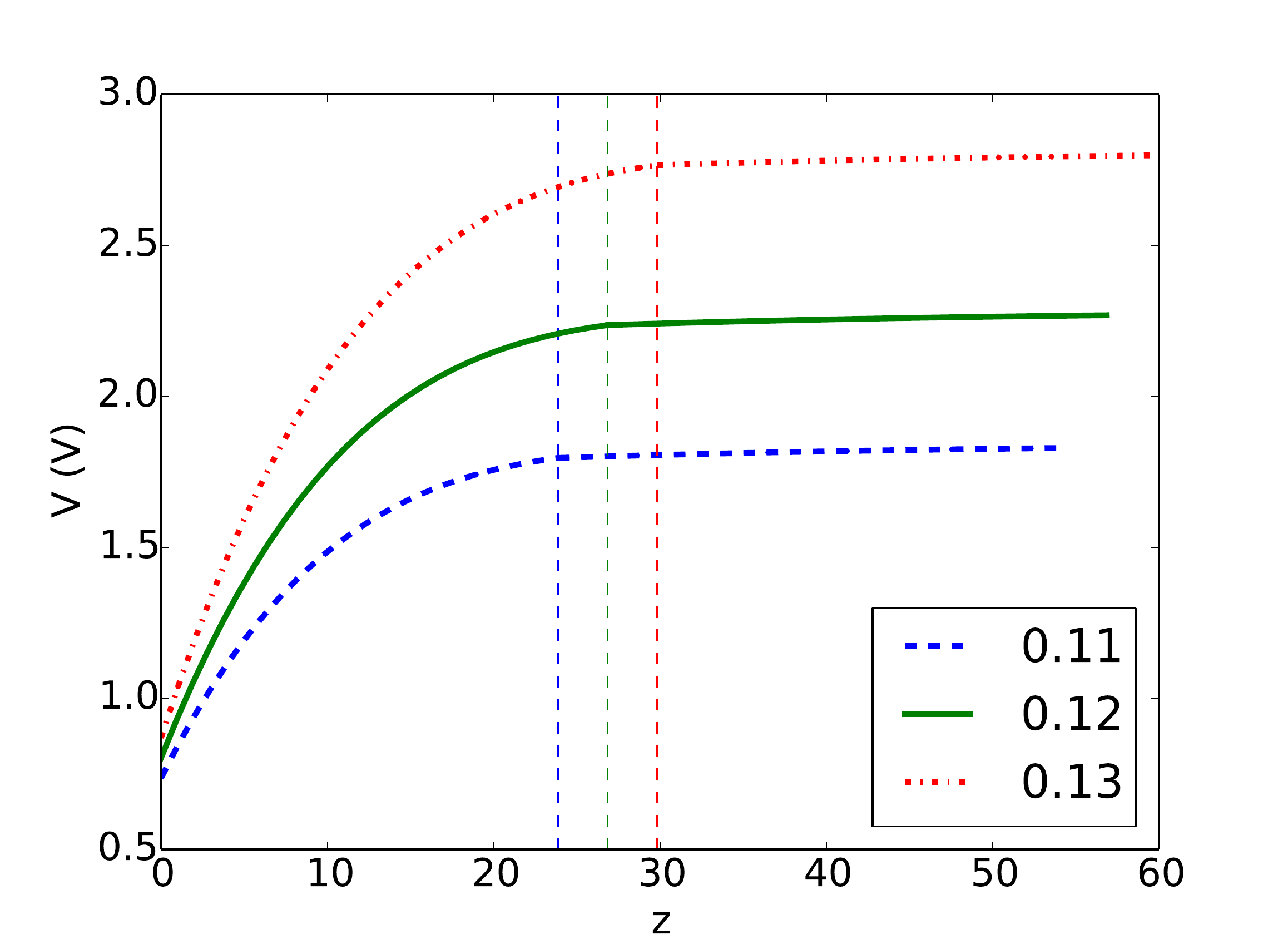}
\caption{\label{V_z} The potential as a function of depth in STO, in the unit of STO lattice constant. The dashed lines represent the boundary of the depletion and the partial depletion regions. Color lines denote different transferred charge density: 0.13 (upper curve), 0.12 (middle) and 0.11 (lowest) e/Fe. }
\end{center}
\end{figure}

So for $z<z_0$, assuming a constant $\epsilon$, 
$\Delta V(z) = c \exp{(z-z_0)/\lambda_\text{TF}}$ where $\lambda_\text{TF} = \sqrt{\epsilon W/8\pi N_de^2}$ is basically the Thomas-Fermi screening length. With the boundary condition that $e  V(z_0) = W/2$, 
\begin{equation}
V(z) = \frac{W}{2e} e^{(z-z_0)/\lambda_\text{TF}}.
\end{equation}
The impurity band has a small band width which we assume as 0.1 eV.
The main potential drop occurs in the depletion region.

Let us assume that for the partial depletion region the dielectric constant is a large constant $\epsilon\approx \epsilon_0 = 500$. This gives us a upper limits of the screening length and the charge that are transferred out. 
\begin{equation}
N_q= N_d\lambda_\text{TF}. 
\end{equation}
With $\lambda_\text{TF}= \sqrt{\epsilon W / 8\pi N_d e^2}$. For the constant $\epsilon$, the $\lambda_\text{TF} = 35.5$\AA\, and $N_q=0.03/$Fe are also both constants.
This means that the depletion region is reduced by, to the upper limit, 35.5\AA.   The potential profile with depth is shown in Fig.~\ref{V_z}. For 0.12 e/Fe surface charge density, the depletion region is now about 27 unit cells. Most of the work function difference should be compensated within this region, leaving 0.05 eV compensated by the partial depleted region where the potential energy decays exponentially.

\bibliography{charge_trans}

\begin{thebibliography}{35}
\bibitem [{lam()}]{lambda-notes}%
  \BibitemOpen
  \href@noop {} {}\bibinfo {howpublished} {We evaluate the two dimensional
  integral of $\lambda$ that is directly from the expression of the self energy
  in Eq.~\ref{Sigma_calc}. A converged value $\lambda\approx 0.5$, is obtained.
  The slightly larger value than our value in the main text comes from the
  contribution where $k^\prime \not\approx k_F$.}\BibitemShut {Stop}%
\end{thebibliography}
\end{document}